\documentclass[twocolumn,showpacs,aps,floatfix]{revtex4}
\usepackage{amsmath,amssymb,eucal,graphicx,bm}
\begin{document}
\title{On the Mixing of Diffusing Particles} 
\author{E.~Ben-Naim}
\affiliation{Theoretical Division and Center for Nonlinear
Studies, Los Alamos National Laboratory, Los Alamos, New Mexico
87545}
\begin{abstract}
We study how the order of $N$ independent random walks in one
dimension evolves with time. Our focus is statistical properties of
the inversion number $m$, defined as the number of pairs that are out
of sort with respect to the initial configuration.  In the
steady-state, the distribution of the inversion number is Gaussian
with the average $\langle m\rangle \simeq N^2/4$ and the standard
deviation $\sigma\simeq N^{3/2}/6$.  The survival probability,
$S_m(t)$, which measures the likelihood that the inversion number
remains below $m$ until time $t$, decays algebraically in the
long-time limit, $S_m\sim t^{-\beta_m}$. Interestingly, there is a
spectrum of $N(N-1)/2$ distinct exponents $\beta_m(N)$. We also find
that the kinetics of first-passage in a circular cone provides a good
approximation for these exponents.  When $N$ is large, the
first-passage exponents are a universal function of a single scaling
variable, $\beta_m(N)\to \beta(z)$ with \hbox{$z=(m-\langle
m\rangle)/\sigma$}.  In the cone approximation, the scaling function
is a root of a transcendental equation involving the parabolic
cylinder equation, $D_{2\beta}(-z)=0$, and surprisingly, numerical
simulations show this prediction to be exact.
\end{abstract}
\maketitle

\section{Introduction}

Consider the permutation $3142$ of the four elements $\{1,2,3,4\}$.
Three pairs: $(1,3)$, $(2,3)$, and $(2,4)$ are inverted in this
permutation.  The inversion number, defined as the total number of
pairs that are out of sort, provides a natural measure for how
``scrambled'' a list of elements is. This basic combinatorial quantity
\cite{pam,wf,gea,mb} is helpful in many contexts. In computer science,
the inversion number plays an important role in sorting and ranking
algorithms \cite{dek}. Common on the web (``customers who like
$\ldots$ may also like $\ldots$''), recommendations for books, songs,
and movies use inversions to quantify how close the preferences of two
customers are \cite{kt}.

\begin{figure}[t]
\includegraphics[width=0.3\textwidth]{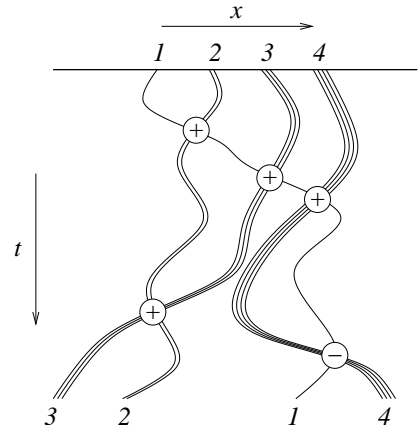}
\caption{Space-time diagram of a four-particle system. The circled
  $+$s and $-$s indicate whether the inversion number increases or
  decreases when two trajectories cross.  Four out of the five
  crossings increase the inversion number, and accordingly, the
  inversion number increases from $m=0$ to $m=3$.}
\label{fig-xt}
\end{figure}

The number of inversions can also be used to measure how the order of
a group of particles in one dimension changes with time.  Figure
\ref{fig-xt} illustrates a space-time diagram of four diffusing
particles. The number of inversions changes whenever two trajectories
cross.  Depending on the initial order of the two respective
particles, a crossing may either introduce a new inversion or undo an
existing one.  Consequently, the inversion number either increases or
decreases by one. Therefore, the inversion number equals the
difference between the number of crossings of the first kind and the
number of crossings of the second kind.

Mixing dynamics has been extensively studied in the context of fluids 
\cite{jmo,vhg} and granular materials \cite{ok}, but much less
attention has been given to mixing in the context of diffusion
\cite{ya,yal}.  In this study, we consider an ensemble of $N$
diffusing particles in one-dimension, a system that is widely used to
model the transport of colloidal and biological particles in narrow
channels \cite{wbl,cdl}.  We use the inversion number to measure the
degree to which particles mix.  Clearly, a persistent small inversion
number indicates a poorly mixed system, while a large inversion number
implies that the opposite is true.

We first study how the distribution of the inversion number evolves
with time. We find that there is a transient regime in which the
average inversion number as well as the standard deviation in this
quantity both grow as the square-root of time.  The distribution of
the inversion number is stationary beyond this transient regime.  When
the number of particles is sufficiently large, the probability
distribution function is always Gaussian, whether in the transient
regime or in the steady-state.

Our main focus is first-passage properties \cite{sr} of the inversion
number. We ask: what is $S_m(t)$, the probability that the inversion
number remains smaller than $m$ up to time $t$. For small values of
$m$, the survival probability $S_m$ measures the likelihood that the
particles remain poorly mixed throughout the evolution. Generally, the
probabilities $S_m$ decay as a power law at large times, $S_m\sim
t^{-\beta_m}$.  In general, there is a broad spectrum of $N(N-1)/2$
distinct exponents,
\hbox{$\{\beta_1,\beta_2,\cdots,\beta_{N(N-1)/2}\}$}, that governs the
asymptotic decay of the survival probabilities.

We heavily use first-passage kinetics of a single particle that
diffuses inside a circular cone \cite{rdd,bs1,bk} to understand the
asymptotic behavior of $S_m$. We first utilize two-dimensional cones
to obtain the first-passage exponents for a three-particle system
exactly. Furthermore, we employ circular cones in $N-1$ dimensions and
find good approximate values for the first-passage exponents.

The cone approximation correctly predicts that when the number of
particles is large, the exponents become a universal function,
$\beta_m(N)\to \beta(z)$, of the scaling variable $z=(m-\langle
m\rangle)/\sigma$.  Here, $\langle m\rangle$ and $\sigma$ are the
average and standard deviation of the distribution of inversion
number, at the steady-state.  Interestingly, our numerical simulations
show that the cone approximation yields the exact scaling function
$\beta(z)$ as a root of a transcendental equation involving the
parabolic cylinder function.

The rest of this paper is organized as follows. In section II, we
introduce our basic system and define the inversion number. Stationary
and transient properties of the distribution of the inversion number
are discussed in sections III and IV, respectively. In section V, we
use the cone approximation to understand first-passage properties of
the inversion number.  Scaling and extremal properties of the
first-passage exponents are the focus of section VI. We conclude in
section VII.

\section{The Inversion Number}

Our goal is to characterize how the order of an ensemble of diffusing
particles changes with time. We conveniently use an ordinary random
walk \cite{ghw,hcb,rg} to model the trajectory of a diffusing particle
\cite{bdup}. Our system includes $N$ identical particles that move on
an unbounded one-dimensional lattice. The particles are completely
independent: at each step one particle is selected at random and it
moves, with an equal probability, either to the left, $x\to x-1$, or
to the right, $x\to x+1$. After each elementary step, time is
augmented by the inverse number of particles, $t\to t+1/N$, so that
each particle moves once per unit time.

We index the particles according to their initial position with the
leftmost particle labeled $n=1$ and the rightmost particle labeled
$n=N$ (Figure \ref{fig-xt}). Let $x_n(t)$ be the position of the $n$th
particle at time $t$. By definition,
\begin{equation}
\label{initial}
x_1(0)<x_2(0)<\cdots <x_{N-1}(0)<x_N(0),
\end{equation} 
but the initial order unravels with time. Consider, for example, the
four-particle system illustrated in Figure \ref{fig-xt}. The particles
reach a state where \hbox{$x_3<x_2<x_1<x_4$} with three pairs,
$(1,2)$, $(1,3)$, and $(2,3)$ being out of sort compared with time
$t=0$.  In general, a pair of particles for which $x_i(t)>x_j(t)$ and
$i<j$ constitutes an inversion. Formally, the total number of
inversions, $m$, is given by 
\begin{equation}
\label{inversion}
m(t)=\sum_{i=1}^N\sum_{j=i+1}^N \Theta\big(x_i(t)-x_j(t)\big).
\end{equation}
Here, $\Theta(x)$ is the Heaviside step function: $\Theta(x)=1$ for
$x>0$ and $\Theta(x)=0$ for $x\leq 0$.  The total number of pairs is
$M=\binom{N}{2}$, and hence, the variable $m$ is within the bounds
$0\leq m \leq M$ with
\begin{equation}
\label{bounds}
M=\frac{N(N-1)}{2}.
\end{equation}
The inversion number is minimal, $m=0$, when the order is exactly the
same as in the initial configuration, and it is maximal, $m=M$, when
the order is the mirror image of the initial state.

The inversion number changes whenever two trajectories cross (Figure
\ref{fig-xt}). A crossing either adds a new inversion or removes an
existing one. Thus, we may assign a positive or a negative ``charge''
to each crossing as illustrated in Figure \ref{fig-xt}. The inversion
number, $m(t)$, is simply the sum of all of the charges up to time
$t$.

\section{The Mahonian Distribution}

We first discuss basic statistical characteristics of the inversion
number including the average, the variance, and more generally, the
probability distribution function.  At large time $t$, each random
walk explores a region of size $\sqrt{t}$, and the probability of
finding the particle at any lattice site inside this region is
effectively uniform. This simple fact already implies that memory of
the initial position fades with time. We thus expect that after
sufficient time elapses, there is no memory of the initial order, and
the order of the particles is completely random.

To understand statistics of the inversion number for randomly ordered
particles we consider the set of all $N!$ permutations of the $N$
elements $\{1,2,\ldots,N\}$. In the random state, each permutation of
these elements occurs with probability $1/N!$. The probability
$P_m(N)$ that the inversion number equals $m$ for a random permutation
is well known as the Mahonian distribution in probability theory
\cite{pam,dek,wf}. We highlight key features of this probability
distribution as it plays a central role in our study.

Let $Q_m(N)=N!P_m(N)$ be the number of permutations of $N$ elements
with exactly $m$ inversions. For example, when $N=3$, one permutation
($123$) is free of inversions, there are two permutations with one
inversion ($213$, $132$), two permutations with two inversions ($312$,
$231$), and a single permutation with three inversions ($321$). Hence,
$Q_0(3)=Q_3(3)=1$ while $Q_1(3)=Q_2(3)=2$. We list the distribution of
the inversion number for $N\leq 4$,
\begin{equation*}
(P_0,P_1,\ldots,P_M)=\frac{1}{N!}\times
\begin{cases}
(1) & N=1,\\
(1,1) & N=2,\\
(1,2,2,1)& N=3,\\
(1,3,5,6,5,3,1) &N=4.
\end{cases}
\end{equation*}
Of course, $P_m(N)$ is nonzero if and only if $0\leq m\leq M$.

Since the mirror image of a configuration with $m$ inversions
necessarily has $M-m$ inversions, the probability distribution
satisfies $P_m=P_{M-m}$. Hence, the distribution is symmetric about
$m=M/2$, and the average $\langle m\rangle\equiv \sum_m m P_m$ is
simply 
\begin{equation}
\label{average}
\langle m\rangle= \frac{N(N-1)}{4}.
\end{equation}
Therefore, the average grows quadratically with the total number of
particles when $N\gg 1$.

The Mahonian distribution satisfies the simple recursion relation
\begin{equation}
\label{recursion}
P_m(N)=\frac{1}{N}\sum_{l=0}^{N-1}P_{m-l}(N-1),
\end{equation}
with $P_m(1)=\delta_{m,0}$.  This recursion reflects that every
permutation of $N$ elements can be generated from a permutation of
$N-1$ elements by inserting the $N$th element in any of the $N$
possible positions. Depending on where this last element is added, the
number of inversions increases by an amount $\Delta m=0,1,\ldots,N-1$.

Let us now introduce the generating function, 
\begin{equation}
\label{generating-def}
{\cal P}(s,N)=\sum_{m=0}^M P_m(N)s^m.
\end{equation}
For instance, ${\cal P}(s,1)=1$, ${\cal P}(s,2)=(1+s)/2!$ and ${\cal
P}(s,3)=(1+s)(1+s+s^2)/3!$. In general, the generating function is
given by the product \cite{dek}
\begin{equation}
\label{generating}
{\cal P}(s,N)=\frac{1}{N!}\prod_{n=1}^N (1+s+s^2+\cdots + s^{n-1}),
\end{equation}
as also follows from the recursion \eqref{recursion}.

We can confirm the average \eqref{average} by differentiating the
generating function once, ${\cal P}'(s=1)=\langle m\rangle$, where the
prime represents differentiation with respect $s$.  By differentiating
the generating function twice and using ${\cal P}''(s=1)=\langle
m(m-1)\rangle$, we obtain the variance \cite{dek}, $\sigma^2=\langle
m^2\rangle -\langle m\rangle^2$,
\begin{equation}
\label{variance}
\sigma^2=\frac{N(N-1)(2N+5)}{72}.
\end{equation}
This expression is obtained from $\sigma^2=\sum_{l=1}^N
\frac{l^2-1}{12}$.  Therefore, the standard deviation is rather large,
$\sigma\simeq N^{3/2}/6$, when $N\gg 1$.

The mean \eqref{average} and the variance \eqref{variance} fully
specify the probability distribution function for an asymptotically
large number of particles. The Mahonian distribution becomes a
function of a single variable, $P_m(N)\to \Phi(z)$, with the scaling
variable
\begin{equation}
\label{scaling-def}
z=\frac{m-\langle m\rangle}{\sigma}.
\end{equation} 
The probability distribution function, $\Phi(z)$, is normal, that is,
a Gaussian with zero mean and unit variance \cite{wf,cjz},
\begin{equation}
\label{normal}
\Phi(z)\simeq\frac{1}{\sqrt{2\pi}}\exp\left(-\frac{z^2}{2}\right).
\end{equation}
To see that the central limit theorem applies, we convert the
generating function into a Fourier transform, and then show that the
Fourier transform is Gaussian in the large-$N$ limit \cite{krb}.

The variable $z$ is a more transparent measure in the following sense.
A value of $z$ of order one implies fairly random order. Indeed,
according to the normal distribution \eqref{normal}, the inversion
number falls within three standard deviations from the mean, $|z|<3$,
with probability $0.997$.  A large value, $|z|\gg 1$, indicates that
the particle order strongly resembles the initial configuration (if
$z>0$) or its mirror image (if $z<0$).

\section{Transient Behavior}

By definition, the inversion number is zero initially, $m(0)=0$.  At
least partially, the initial order is preserved in the early stages of
evolution, and the number of inversions must be substantially lower
than \eqref{average}.

We consider the natural initial condition where the particles occupy
$N$ consecutive lattice sites: $x_i(0)=i$, for all
$i=1,2,\ldots,N$. Early on, particles ``interact'' only within their
local neighborhood.  The interaction length, $\ell$, grows diffusively
with time, $\ell\sim \sqrt{t}$.  On this length scale, particles are
well-mixed, and according to \eqref{average}, the number of inversions
{\em per particle} is proportional to the number of interacting
particles, $\ell$. Hence, the average number of inversions grows
according to $\langle m(t)\rangle\sim N\ell \sim N\sqrt{t}$. As a
consequence,
\begin{equation} 
\label{mt}
\langle m(t)\rangle\simeq  
\begin{cases}
{\rm const.}\times N\sqrt{t}\quad & 1\ll t\ll N^2,\\
N^2/4\quad & N^2\ll t,
\end{cases}
\end{equation}
when $N\gg 1$.  The two expressions match at \hbox{$t\sim N^2$}, a
diffusive time scale that can be viewed as the mixing time.  Therefore,
there is a transient regime in which the inversion number grows as the
square-root of time, followed by a steady-state, in which the average
is given by \eqref{average}.

We obtain the variance using a similar heuristic argument. According
to \eqref{variance}, the variance per particle is quadratic in the
number of interacting particles, \hbox{$\sigma^2\sim N
\ell^2$}. Therefore, \hbox{$\sigma^2\sim N\,t$} in the transient
regime, 
\begin{equation} 
\label{sigmat}
\sigma(t)
\simeq  
\begin{cases}
{\rm const}\times \sqrt{N\,t}\quad & 1\ll t\ll N^2,\\
N^{3/2}/6\quad & N^2\ll t.
\end{cases}
\end{equation}
As expected, the transient behavior matches the steady-state behavior
\eqref{variance} at the diffusive time scale $t\sim N^2$. Like the
average, the standard deviation also grows as the square root of time.

\begin{figure}[t]
\includegraphics[width=0.43\textwidth]{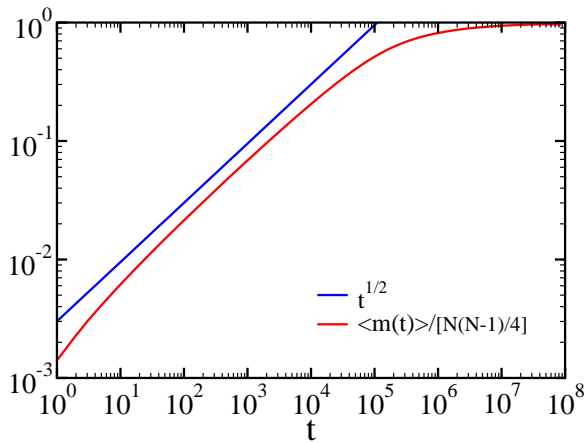}
\caption{The normalized average $\langle m(t)\rangle/[N(N-1)/4]$
  versus time $t$. The results correspond to an average over $10^2$
  independent realizations of a system with $N=10^3$ random
  walks. Also shown for reference is a line with slope $1/2$.}
\label{fig-average}
\end{figure}

As shown in Figure \ref{fig-average}, results of numerical simulations
confirm the scaling behavior \eqref{mt}. Moreover, the numerically
measured average matches the steady-state value corresponding to the
Mahonian distribution. We also verified that the stationary
distribution is Gaussian with the variance \eqref{variance}.

\begin{figure}[t]
\includegraphics[width=0.44\textwidth]{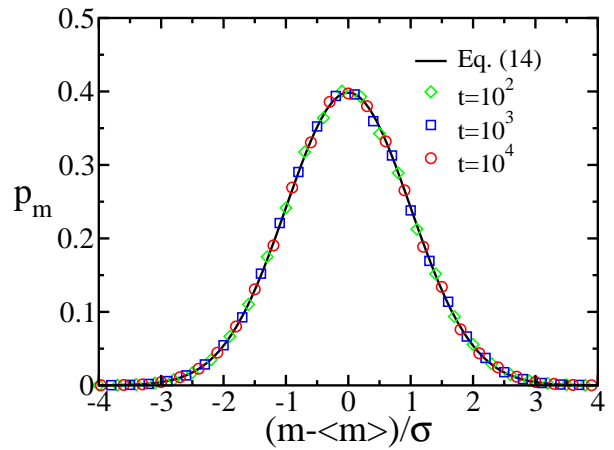}
\caption{The distribution of the inversion number in the intermediate
time regime $1\ll t\ll N^2$. Shown is the distribution $p_m\equiv
p_m(N,t)$ versus the variable $\big(m-\langle
m(t)\rangle\big)/\sigma(t)$.  The results are from $10^5$ independent
realizations of a system with $N=10^3$ random walks. The distribution
is shown at times $t=10^2$ (diamonds), $t=10^3$ (squares), and
$t=10^4$ (circles).  Also shown for reference is the normal
distribution \eqref{normal1}.}
\label{fig-distribution}
\end{figure}

The simulations also show that the time-dependent distribution of
inversion number, $p_m(N,t)$, is Gaussian throughout the transient
regime (Figure \ref{fig-distribution}):
\begin{equation}
\label{normal1}
p_m(N,t)\simeq
\frac{1}{\sqrt{2\pi\sigma^2(t)}}\exp\left[-\frac{(m-\langle
m(t)\rangle)^2}{2\sigma^2(t)}\right].
\end{equation}
This behavior provides further support for our heuristic
argument. Indeed, if the particles are well-mixed locally, then the
distribution of the number of inversions per particle is Gaussian, and
as the sum of $N$ Gaussian variables, the total inversion number must
also have a Gaussian distribution.

We used two different algorithms to simulate the diffusion process.
In the naive algorithm, we randomly select a particle and move it to a
randomly-chosen neighboring site. We increase time by $1/N$ after each
jump. To calculate the inversion number, we  use the formula
\eqref{inversion}, but since this enumeration requires ${\cal O}(N^2)$
operations, this simulation method is inefficient at large $N$.

To overcome this difficulty, we introduced a variant where each
lattice site may be occupied by at most one particle.  At each step we
pick one particle at random and attempt to move it by one site. This
move is always accepted if the neighboring site is vacant, but
otherwise, it is accepted with probability $1/2$. In the latter case,
we merely exchange the identities of the respective particles, and as
appropriate, update the inversion number by either $+1$ or $-1$.  In
our implementation, there are two arrays: the first lists the particle
positions, {\em in order}, and the second lists the original position
of each particle in the first list.  This algorithm has a fixed
computational cost per step, and it automatically keeps track of the
inversion number.  We rely on the fact that in one dimension,
noninteracting random walks are equivalent to random walks that
interact by exclusion \cite{teh,dgl,ra,bs}. Still, we verified that
the two algorithms yield essentially the same results.  We utilized
the naive algorithm to simulate small systems with $N<10$, but
otherwise, we used the efficient algorithm.

\section{First-Passage Kinetics}

We have seen that the inversion number, which grows quadratically with
the number of particles, can be quite large.  Yet, if the mixing is
poor and the particle trajectories rarely cross, the inversion number
remains small.  To quantify how common such a scenario is, we study
first-passage kinetics \cite{sr,snm}. In particular, we ask: what is
the probability, $S_m(t)$, that the inversion number remains smaller
than $m$ until time $t$. This ``survival'' probability is closely
related to the first-passage probability as $[-dS_m/dt]\times dt$ is
the probability that the inversion number reaches $m$ for the first
time during the infinitesimal time interval $(t,t+dt)$.

The quantity $S_1$ is the probability that the original order is
perfectly maintained, or equivalently, the likelihood that none of the
trajectories cross.  This survival probability decays as a power law,
with a rather large exponent,
 \begin{equation}
\label{first}
S_1\sim t^{-N(N-1)/4},
\end{equation}
in the long-time limit \cite{mef,hf,fg,gz,djg,ck}.  Our goal is to
understand how this asymptotic behavior changes as the threshold $m$
increases.

When $N=2$, the separation between the two random walks itself
undergoes a random walk. Hence, $S_1$ is equivalent to the survival
probability of a one-dimensional random walk in the vicinity of a
trap, and $S_1\sim t^{-1/2}$ in agreement with \eqref{first}.

When $N=3$, we conveniently map the three random walks onto a single
``compound'' random walk in three dimensions with the coordinates
$(x_1,x_2,x_3)$. To find $S_1$, we require that the compound walk
remains inside the region $x_1<x_2<x_3$. We may view the boundary of
this region as absorbing, and then, $S_1(t)$ equals the likelihood
that that the compound walk survives at time $t$.  The absorbing
boundary forms a wedge because it is defined by the intersection of
two planes, $x_1=x_2$ and \hbox{$x_2=x_3$}.  Generally, the survival
probability of a particle that diffuses inside an absorbing wedge
decays algebraically,
\begin{equation}
S\sim t^{-1/(4V)},
\end{equation}
where $V=\alpha/\pi$ is the normalized opening angle \cite{fs}.  (The
opening angle $0<\alpha\leq \pi$ is the angle between the wedge axis
and the wedge boundary.)  Alternatively, $0<V\leq 1$ is the fraction
of the total solid angle enclosed by the wedge. The region
$x_1<x_2<x_3$ occupies a fraction $V_1=\frac{1}{3!}=\frac{1}{6}$ of
space and hence, $S_1\sim t^{-3/2}$, as also follows from
\eqref{first}. To find $S_2$ and $S_3$, we note that the regions in
which the compound walk is allowed to move are always wedges (the
three planes $x_1=x_2$, $x_1=x_3$, and $x_2=x_3$ divide space into six
equal wedges \cite{bjmkr}.) Moreover, the fraction of total solid
angle enclosed by the absorbing boundaries is given by the cumulative
distribution of inversion number: \hbox{$V_2=\frac{1}{3!}+
\frac{2}{3!}=\frac{1}{2}$} and \hbox{$V_3=\frac{1}{3!}+
\frac{2}{3!}+\frac{2}{3!}=\frac{5}{6}$}.  Hence, all three survival
probabilities decay algebraically with time \cite{sr},
\begin{equation}
\label{three}
S_1\sim t^{-3/2},\qquad S_2\sim t^{-1/2},\qquad S_3\sim t^{-3/10},
\end{equation}
and there are three distinct first-passage exponents.

\begin{figure}[t]
\includegraphics[width=0.43\textwidth]{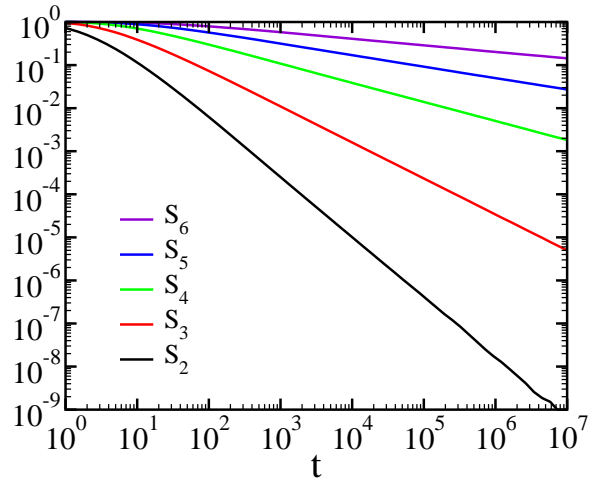}
\caption{The survival probability $S_m(t)$ versus $t$ for a
  four-particle system. Shown are the quantities $S_2$ (bottom curve),
  $S_3$, $\ldots$, $S_6$ (top curve). The number of independent
  Monte Carlo runs varies from $10^6$ for the slowest decay to
  $10^{12}$ for the fastest decay.}
\label{fig-beta4}
\end{figure}

The asymptotic behaviors \eqref{three} suggest that all of the
survival probabilities decay algebraically, 
\begin{equation}
\label{decay}
S_m\sim t^{-\beta_m},
\end{equation}
in the long-time limit. Moreover, there is a large family of 
exponents
\begin{equation}
\label{family}
\beta_1>\beta_2>\cdots>\beta_{N(N-1)/2},
\end{equation}
that characterizes the power-law decay \eqref{decay}.  We stress that
the exponents depend on two variables, the threshold $m$ and the
number of particles $N$, \hbox{$\beta\equiv \beta_m(N)$}.  We already
know the exact values $\beta_1(3)=3/2$, $\beta_2(3)=1/2$, and
$\beta_3(3)=3/10$ as well as $\beta_1(N)=N(N-1)/4$.

Our numerical simulations confirm that indeed, there is a large
spectrum of exponents. As shown in Figure \ref{fig-beta4}, there are
six decay exponents when $N=4$. Table I lists the numerically measured
values $\beta_m$, obtained from the local slope $d\ln S_m/d\ln t$.

In general, the compound walk is confined to a certain ``allowed''
region of space. This region is bounded by multiple intersecting
planes of the type $x_i=x_j$ with $i\neq j$, and generally, this
unbounded domain has a complicated geometry.  The boundary of this
region encloses a fraction $V_m(N)$ of the total solid angle. On
combinatorial grounds alone, we conveniently deduce that this
fraction is given by the cumulative Mahonian distribution
\begin{equation}
\label{cumulative-def}
V_m(N)=\sum_{l=0}^{m-1} P_l(N).
\end{equation}
Since the Mahonian distribution is symmetric, we have
$V_m+V_{M+1-m}=1$. To evaluate $V_m(N)$, we expand the generating
function \eqref{generating}, and for $m\leq 4$, we have \cite{dek}
\begin{equation}
V_m(N)\!=\!\frac{1}{N!}\!\times\! 
\begin{cases}
1 & m=1,\\
N & m=2,\\
\frac{1}{2}(N-1)(N+2) & m=3,\\
\frac{1}{6}(N+1)(N^2+2N-6) & m=4.\\
\end{cases}
\end{equation}

We have seen that the allowed region is a wedge when $N=3$. To obtain
an approximation for the first-passage exponents, we follow an
approach that proved useful in other first-passage problems involving
multiple random walks and replace the boundary of the allowed region
with a suitably chosen cone in $N-1$ dimensions \cite{bk1}.  An
unbounded cone with opening angle $\alpha$ occupies a fraction
$V(\alpha)$ of the total solid angle, given by
\begin{equation}
\label{Valpha}
V(\alpha)=\frac{\int_0^\alpha d\theta\,(\sin\theta)^{N-3}} {\int_0^\pi
d\theta\,(\sin\theta)^{N-3}}.
\end{equation}
In $d$ dimensions, we have $d\Omega\propto (\sin\theta)^{d-2}d\theta$
where $\Omega$ is the solid angle and $\theta$ is the polar angle in 
spherical coordinates.  In the cone approximation, we require
\begin{equation}
\label{Valpha-equal}
V(\alpha)=V_m
\end{equation}
with $V_m$ given in \eqref{cumulative-def}. 

\begin{table}[t]
\begin{tabular}{|c|c|c|c|c|c|c|}
\hline $m$&$1$&$2$&$3$&$4$&$5$&$6$\\ 
\hline 
$V_m$&$\frac{1}{24}$& $\frac{1}{6}$& $\frac{3}{8}$& $\frac{5}{8}$& $\frac{5}{6}$ & $\frac{23}{24}$\\ 
[2pt]\hline 
$\alpha_m$ & $0.41113$ & $0.84106$ & $1.31811$ & $1.82347$ & $2.30052$ & $2.73045$ \\
\hline
$\beta_m^{\rm cone}$& $2.67100$ & $1.17208$ & $0.64975$ & $0.39047$ & $0.24517$ & $0.14988$  \\
\hline $\beta_m$& $3$ & $1.39$ & $0.839$ & $0.455$ & $0.275$ & $0.160$ \\ 
[2pt] \hline
\end{tabular}
\caption{The six first-passage exponents for a four-particle
system. The values $\beta_m$ are from the Monte Carlo simulation results
shown in Figure \ref{fig-beta4}. The values $\beta_m^{\rm cone}$ were
obtained using the cone approximation, specified in
Eqs.~\eqref{cumulative-def}-\eqref{cone}. The cumulative Mahonian
distribution, $V_m$, and the opening angle, $\alpha_m$, are listed as
well.}
\end{table}

In a cone, the first-passage exponent $\beta\equiv \beta(\alpha)$
decreases as the opening angle $\alpha$ increases.  In particular,
$\beta=\pi/4\alpha$ in two dimensions, and
$\beta=(\pi-\alpha)/2\alpha$ in four dimensions.  Generally, however,
$\beta$ is the smallest root of the following transcendental equation
involving the associated Legendre functions \cite{NIST} of degree
$2\beta+\gamma$ and order $\gamma=\frac{N-4}{2}$ \cite{bk}
\begin{eqnarray}
\label{cone}
\begin{split}
Q_{2\beta+\gamma}^\gamma(\cos\alpha) &= 0\qquad N\ {\rm odd},\\
P_{2\beta+\gamma}^\gamma(\cos\alpha) &= 0\qquad N\ {\rm even}.
\end{split}
\end{eqnarray}
Regardless of the dimension, the surface of a cone with $\alpha=\pi/2$
is a plane, and hence, $\beta(\pi/2)=1/2$.

\begin{figure}[t]
\includegraphics[width=0.5\textwidth]{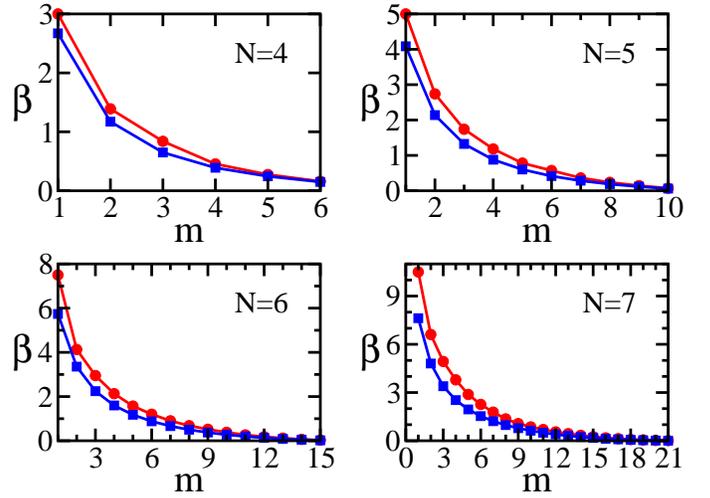}
\caption{The first-passage exponent $\beta_m$ versus $m$ for
$N=4,5,6,7$. Shown are simulation results (circles) and the outcome of
the cone approximation (squares).}
\label{fig-beta5}
\end{figure}

For example, to find $\beta_1(4)$, we first determine the fraction
$V_1(4)=\frac{1}{4!}=\frac{1}{24}$ using \eqref{cumulative-def}. Then,
we calculate the opening angle $\alpha=0.41113$ using equations
\eqref{Valpha}-\eqref{Valpha-equal} and finally determine the exponent
$\beta_1(4)=2.67100$ as the appropriate root of equation \eqref{cone}.
By construction, the cone approximation is exact for three
particles. This approach gives a useful approximation to the six
first-passage exponents when $N=4$ (Table I).  Remarkably, the cone
approximation continues to be a good approximation as the number of
particles increases (Figure \ref{fig-beta5}).

\section{The Scaling Function}

We are especially interested in the behavior when the number of
particles is large.  Let us first evaluate the cumulative Mahonian
distribution in the large-$N$ limit. Since the Mahonian distribution
is normal, the cumulative distribution is given by the error function,
\begin{equation}
\label{Vm-scaling}
V_m(N)\to \frac{1}{2}+\frac{1}{2}\,{\rm erf} \left(\frac{z}{\sqrt{2}}\right),
\end{equation}
when $N\to\infty$. Here, $z$ is the scaling variable defined in
\eqref{scaling-def} and ${\rm erf}(\xi)=(2/\sqrt{\pi})\int_0^\xi
\exp(-u^2)du $. To obtain Eq.~\eqref{Vm-scaling}, we substitute
\eqref{normal} into \eqref{cumulative-def} and convert the sum into an
integral. Equation \eqref{Vm-scaling} is relevant in the limit
$N\to\infty$, $m\to\infty$ with the scaling variable $z$ finite.

Next, we evaluate the solid angle enclosed by an unbounded cone 
when the dimension is large. The dominant contribution to the integral
in \eqref{Valpha} comes from a narrow region of order $1/\sqrt{N}$
centered on $\alpha=\pi/2$ where the integrand is Gaussian,
\begin{equation*}
(\sin\theta)^{N-2} \simeq e^{-N(\pi/2-\theta)^2/2}.
\end{equation*}
Using \hbox{$\int_{-\infty}^\infty
  \exp\big[\!-\!N(\pi/2-\theta)^2/2\big]d\theta\to \sqrt{2\pi/N}$}, we
find that the fraction $V(\alpha)$ has the scaling form
\begin{equation}
\label{Valpha-scaling}
V(\alpha,N)\to \frac{1}{2}+\frac{1}{2}\,{\rm erf}\left(\frac{-y}{\sqrt{2}}\right), 
\end{equation}
with $y=(\cos\alpha)\sqrt{N}$.  In writing this equation, we used the
facts that \hbox{$\cos\alpha\simeq\pi/2-\alpha$} and ${\rm
erf}(\xi)=-{\rm erf}(-\xi)$.  Equation \eqref{Valpha-scaling} holds in
the limit $N\to\infty$, $\alpha\to \pi/2$, with the scaling variable
$y$ finite.

Asymptotic analysis of equation \eqref{cone} shows that the exponent
$\beta(\alpha)$ adheres to the scaling form \cite{bk} 
\begin{equation}
\label{betaalpha-scaling}
\beta(\alpha,N)\to \beta(y)\quad{\rm with}\quad y=(\cos\alpha)\sqrt{N},
\end{equation}
in the limit $N\to\infty$, $\alpha\to\pi/2$ with the scaling variable
$y$ finite.  The scaling function, $\beta(y)$, is specified by the
transcendental equation $D_{2\beta}(y)=0$, where $D_\nu$ is the
parabolic cylinder function of order $\nu$ \cite{NIST}. The smallest 
root is the appropriate one \cite{bk}.

\begin{figure}[t]
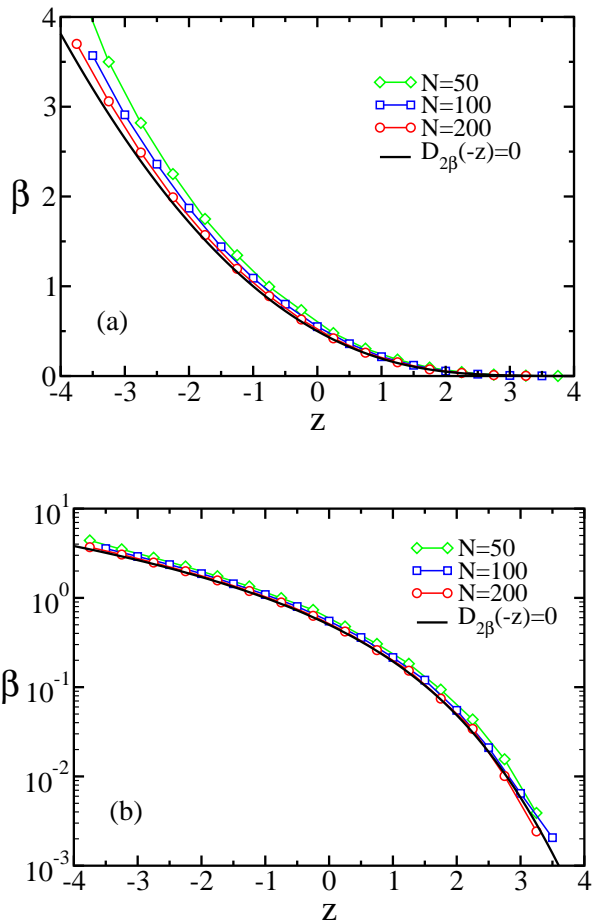

\includegraphics[width=0.425\textwidth]{fig6a}
\includegraphics[width=0.44\textwidth]{fig6b}
\caption{The exponent $\beta$ versus the scaling variable $z$, shown 
using: (a) a linear-linear plot and (b) a linear-log plot.  The simulation
results are from Monte Carlo runs with $N=50$ (diamonds), $N=100$
(squares), and $N=200$ (circles) particles. The number of independent
realizations varies from $10^4$ for slow first-passage processes to
$10^8$ for fast one.  The solid line shows the theoretical prediction
\eqref{scaling-function}.}
\label{fig-beta}
\end{figure}

By comparing equations \eqref{Vm-scaling} and \eqref{Valpha-scaling},
we find our main result: the first-passage exponent depends on a
single scaling variable,
\begin{equation}
\label{scaling-form}
\beta_m(N) \to \beta(z) \quad {\rm with}\quad 
z=\frac{m-\langle m\rangle}{\sigma},
\end{equation}
in the large-$N$ limit. We reiterate that the average $\langle
m\rangle$ and the standard-deviation $\sigma$ correspond to the
steady-state values \eqref{average} and \eqref{variance},
respectively.  Using $y=-z$, the scaling function $\beta(z)$ is 
the smallest root of the transcendental equation
\begin{equation}
\label{scaling-function}
D_{2\beta}(-z)=0,
\end{equation}
involving the parabolic cylinder function.  When $\beta$ is a
half-integer, the parabolic cylinder function is related to the
Hermite polynomials \cite{NIST} and using this equivalence, we have
$\beta(0)=1/2$, $\beta(-1)=1$, and $\beta(-\sqrt{3})=3/2$.

Our numerical simulations (Figure \ref{fig-beta}) confirm that the
exponents $\beta_m(N)$ have the scaling form
\eqref{scaling-form}. Interestingly, the simulations strongly suggest
that the scaling function predicted by the cone approximation is {\em
exact}. We note that the convergence to the infinite-particle limit is
very fast for positive $z$, but much slower for negative $z$
\cite{bk}.

With the power-law decay \eqref{decay}, the mean first-passage time
diverges whenever $\beta<1$, but it is finite otherwise. Since
$\beta(z=-1)=1$, the time required for the inversion number to reach
one standard deviation from the mean is infinite, on average.
Regardless of the threshold $z$, there is a considerable chance that
the random walks are poorly mixed because the survival probability
decays algebraically.

The scaling behavior is remarkable for a number of reasons. First, the
form of the scaling variable, \hbox{$z\equiv (m-N^2/4)/(N^{3/2}/6)$},
is quite unusual. Second, there are roughly $N^2/2$ first-passage
exponents and numerical evaluation of this large spectrum is daunting.
Yet, the scaling form \eqref{scaling-form} gives the range of
parameters for which $\beta$ is of order one, and hence, numerically
measurable.  (It is difficult to measure a vanishing exponent,
$\beta\to 0$, or a divergent exponent, $\beta\to \infty$.)  Last, the
emergence of scaling laws for a family of scaling exponents is also
intriguing. Typically, in Statistical Physics, the opposite is true as 
one or two scaling exponents characterize a scaling law \cite{hes}.

The extremal behaviors of the roots of the transcendental equation
\eqref{scaling-function} are derived in ref.~\cite{bk},
\begin{equation}
\label{scaling-limits}
\beta(z)\sim
\begin{cases}
z^2/8 &z\to-\infty,\\
\sqrt{z^2/8\pi}\exp\left(-z^2/2\right)& z\to\infty.
\end{cases}
\end{equation}
The first-passage exponent is algebraically large if $z$ is large and
negative, but it is exponentially small if $z$ is large and positive.

\begin{table}[t]
\begin{tabular}{|c|c|c|c|c|c|c|}
\hline
$N$&$3$&$4$&$5$&$6$&$7$&$8$\\
\hline
$\beta_1$   & $\frac{3}{2}$ &    $3$      &  $5$        & $\frac{15}{2}$     & $\frac{21}{2}$     & $14$ \\
[1pt]\hline 
$\beta_1^{\rm cone}$ & $\frac{3}{2}$ &  $2.67100$ &  $4.08529$ & $5.73796$ & $7.62336$ & $9.73686$ \\
[1pt]\hline 
$\beta_M^{\rm cone}$ & $\frac{3}{10}$ &  $0.14988$ &  $0.061195$ & $0.019895$ & $0.0050713$ & $0.0010266$ \\
[2pt]
\hline
\end{tabular}
\caption{The largest exponent, $\beta_1^{\rm cone}$, and the smallest 
exponent, $\beta_M^{\rm cone}$, obtained using the cone approximation
for $N\leq 8$.  Also listed for reference, is the exact value 
$\beta_1$.}
\end{table}

The exponential decay in \eqref{scaling-limits} implies that it is
extremely unlikely that the initial order is perfectly reversed.  The
smallest exponent $\beta_M$ characterizes the probability $S_M$ that
the order of the walkers does not turn into the mirror image of the
initial state, that is, the probability that the compound walk remains
in the {\em exterior} of the so-called ``Weyl chamber''
$x_1<x_2<\cdots<x_N$ \cite{mef,hf,fg,gz,djg,ck}. This domain has
$V_M=1-\frac{1}{N!}$, and Table II lists the outcome of the cone
approximation for small $N$.  To find the outcome of the cone
approximation at large $N$, we first estimate the opening angle,
$\pi-\alpha\simeq e/N$ by using Eq.~\eqref{Valpha} and the Stirling
formula $N!\simeq \sqrt{2\pi N}(N/e)^N$. From the asymptotic behavior
for wide cones at large dimensions, \hbox{$\beta\simeq
\sqrt{N/8\pi}(\pi-\alpha)^{N-3}$} \cite{bk}, we conclude \cite{bk1}
\begin{equation}
\label{smallest}
\beta_M\simeq \frac{N^4}{2\,e^3\,N!}.
\end{equation}
This value is extremely small, decaying roughly as the inverse of a
factorial, and it is impossible to measure such a minuscule quantity
using numerical simulations.

The largest exponent describes the probability that the particles
maintain the initial order or that the compound walk remains in the
{\em interior} of the Weyl chamber with $V_1=\frac{1}{N!}$. Table II
compares the outcome of the cone approximation with the exact value
$\beta_1=N(N-1)/4$. The quality of the cone approximation worsens as
$N$ grows. Nevertheless, the cone approximation is qualitatively
correct. By substituting the opening angle $\alpha\simeq e/N$ into the
thin-cone asymptotic behavior $\beta(\alpha)\simeq N\alpha^{-1}/4$
\cite{bk}, we find $\beta_1\simeq N^2/(4e)$. This expression captures
the quadratic growth of the exponent. Remarkably, the cone
approximation is exact inside the scaling region, but it is only
approximate outside this region.

\section{Conclusions}

In summary, we used the number of pair inversions to measure the
one-dimensional mixing of independent diffusing trajectories. A high
inversion number typifies strong mixing whereas a persistent small
inversion number indicates poor mixing.  In the steady-state, the
distribution of inversion number is given by the well-known Mahonian
distribution, and consequently, it is Gaussian when the number of
particles is large. Preceding the steady-state is a transient regime
in which both the average inversion number and the standard deviation
grow diffusively with time.

We focused on first-passage statistics and showed that the probability
that the inversion number does not exceed a certain threshold decays as
power law with time. Moreover, we found that a large spectrum of decay
exponents characterizes the asymptotic behavior. When the number of
particles is large, the exponents obey a universal scaling
function. The scaling variable equals the distance between the
threshold inversion number and the average inversion number, measured
in terms of the standard deviation.

The cone approximation, which replaces the region in which the
compound random walk is allowed to move with an unbounded circular
cone, plays a central role in our analysis.  This approach is exact
for three particles, it produces very good estimates in
higher dimensions, and remarkably, this framework yields the exact
scaling function. The cone approximation gives lower bounds for the
decay exponents because, among all unbounded domains with the same
solid angle, the perfectly circular cone maximizes the survival
probability \cite{bk1,jwsr,ch}. The cone approximation is useful in
answering other first-passage questions such as the probability that
the $n$th rightmost random walk does not cross the origin and the
probability that the original rightmost particle always remains ahead
of at least $n$ other particles \cite{bk1}. In both cases, there are
as many exponents as there are particles, and curiously, the circular
cone framework produces the scaling function governing the
first-passage exponents approximately in the first case and exactly in
the second case.

Understanding when the cone approximation is exact and when it is
approximate is an interesting challenge, with implications well beyond
first-passage \cite{cj,jdj,ntv}. The first-passage exponent is
directly related to the lowest eigenvalue of the Laplace operator, and
therefore, we conclude that the lowest eigenvalue of the Laplacian
similarly obeys scaling laws in high dimensions. The shape of the
scaling function depends on the actual geometry \cite{ic}.

\medskip
I thank Paul Krapivsky and Timothy Wallstrom for
useful discussions. This research is supported by DOE grant
DE-AC52-06NA25396.

\end{document}